\documentclass[prl,twocolumn,showpacs,preprintnumbers,amsmath,amssymb, superscriptaddress]{revtex4-1}

\usepackage[makeroom]{cancel}

\usepackage{amsmath}    % need for subequations
\usepackage{amssymb}
\usepackage{graphicx}   % need for figures
\usepackage{verbatim}   % useful for program listings
\usepackage{color}      % use if color is used in text
\usepackage{hyperref}   % use for hypertext links, including those to external documents and URLs
\usepackage[normalem]{ulem}
\usepackage{natbib}
\usepackage{fixmath}
\usepackage{enumitem}

\hypersetup{colorlinks,linkcolor=blue,urlcolor=blue,citecolor=blue}

\definecolor{darkgreen}{rgb}{0,0.5,0}
\definecolor{purple}{rgb}{0.35,0,0.35}
\definecolor{orange}{rgb}{1,0.5,0}
\definecolor{darkred}{rgb}{.7,0,0}
\definecolor{darkblue}{rgb}{0,0,.3}
\definecolor{grey}{rgb}{.6,.6,.6}
\definecolor{dimgreen}{rgb}{0.2,0.6,0.1}
\newcommand{\ket}[1]{|#1\rangle}
\newcommand{\bra}[1]{\langle#1|}
\newcommand{\average}[1]{\langle #1 \rangle}

\newcommand{\cN}{\cal N}

\newcommand{\toq}{\tilde \omega_q}
\newcommand{\oq}{\omega_q}

\begin{document}
\definecolor{darkgreen}{rgb}{0,0.5,0}

\title{Universal conductance of a PT-symmetric Luttinger liquid after a quantum quench}

\author{C\u at\u alin Pa\c scu Moca}
\email{mocap@uoradea.ro}
\affiliation{MTA-BME Quantum Dynamics and Correlations Research Group, Institute of Physics,
Budapest University of Technology and Economics, Budafoki ut 8., H-1111 Budapest, Hungary}
\affiliation{Department  of  Physics,  University  of  Oradea,  410087,  Oradea,  Romania}
\author{Bal\'azs D\'ora}
\affiliation{Department of Theoretical Physics and MTA-BME Lend\"ulet Topology and Correlation Research Group,
Budapest University of Technology and Economics, 1521 Budapest, Hungary}
%\author{Spiros Sotiriadis}
%\affiliation{ University of Ljubljana, Faculty of Mathematics and Physics, Jadranska 19, SI-1000 Ljubljana, Slovenia}

\date{\today}

\begin{abstract}
We study the non-equilibrium dynamics and transport of a  PT-symmetric Luttinger liquid (LL) after an interaction quench.
The system is prepared in domain wall initial state.  After a quantum quench to spatially homogeneous, PT-symmetric LL, the domain wall  
develops into a flat central region that spreads out ballistically faster than the conventional Lieb-Robinson maximal speed.
By evaluating the current inside the regular lightcone,  we find a universal conductance $e^2/h$, insensitive to the strength of the PT-symmetric interaction.
On the other hand, by repeating the very same time evolution with a hermitian LL Hamiltonian, the conductance is heavily renormalized by the hermitian interaction as $e^2/hK$ with $K$ the LL parameter.
Our analytical  results are tested numerically, confirming 
the  universality of the conductance in the non-hermitian realm.
\end{abstract}
\maketitle

\emph{Introduction.}
An important class of systems described by non-Hermitian Hamiltonians has emerged recently in quantum physics. 
Such models exhibit several exciting new phenomena such as exceptional point~\cite{stehmann,Heiss.2004, Miri.2019,Ozdemir.2019}, non-Hermitian skin effect with the majority of eigenstates localized at the boundaries~\cite{Yao.2018,Longhi.2019}, topological transitions~\cite{zeuner,tonylee,zhou18}, anomalous transport behavior~\cite{Eichelkraut.2013,Schomerus.2014,Longhi.2015} to mention a few. 
In general, a non-Hermitian Hamiltonian can arise naturally as the back-action to continuous monitoring and controlled post 
selection measurement that suppress the quantum jump processes~\cite{Ashida.2016}. Examples include inelastic one and two body 
losses in ultracold atomic lattices~\cite{Turlapov.2003,Syassen.2008}. As such systems are under continuous surveillance, their state 
evolves in time under non-equilibrium conditions. 

In this context, a key issue concerns the spreading of correlations
captured by the typical light-cone effect following a quench.  
It has been shown by Lieb and Robinson~\cite{Lieb.1972}
that  in Hermitian systems with short range interactions a maximal velocity for spreading the information does exists. 
On the other hand, 
in the non-Hermitian realm this maximum boundary is exceeded, as higher supersonic modes are developed~\cite{Ashida.2018}, travelling with velocities that are multiples of the regular Lieb-Robinson sound velocity~\cite{Dora.2020}. 

So far, the transport and dynamics following a quantum quench have been studied thoroughly in Hermitian models. 
A variety of different models have been considered, which can be framed into two
main classes: the first consists on non-equilibrium dynamics in spatially  homogeneous systems such as 
in spin$-{1 \over 2}$ XXZ chain~\cite{Pollmann.2013,Mesty.2015,Schoenauer.2019} or  Hubbard chains~\cite{Moeckel.2008,Luchli.2008,Bertini.2017}. In the second class the initial state 
is spatially inhomogeneous. Relevant examples are domain wall structures in XXZ models~\cite{Mitra.2010,Stephan.2017,Misguich.2019} or systems
featuring local impurities~\cite{Schiro.2010,Vasseur.2013,Schwarz.2018}. 
In a previous publication~\cite{Dora.2020}
we have analyzed the behaviour after the quantum quench in a homogeneous PT-symmetric Luttinger liquid (LL), and found that the typical LL 
behaviour is preserved in the long time limit, but at short times, the non-unitary evolution generates supersonic modes. 

Transport in a LL has long been investigated. In a clean LL, the conductance was found to be strongly, renormalized by the interaction~\cite{giamarchi}. However, it was found later that in an ideal LL connected to leads~\cite{safi,maslov}, 
the dc conductance depends only on the properties of the leads of a quantum wire containing a Luttinger liquid, and is given by the conductance quantum, $e^2/h$ per spin orientation, regardless of the interactions in the wire. 

The  purpose of this paper is to extend the previous analysis and investigate  the 
quench dynamics and the universality of transport in an inhomogeneous non-Hermitian Luttinger
liquid initially prepared in a domain wall state~\cite{Das.2019}.
%The system is subject to a sudden quench: At  $t=0$  the chemical potential
%that sustains the domain wall is turned off, and at the same time the exchange 
%interaction of a non-Hermitian type is switched on. 
%We are interested in the complete spatial and time 
%dependence profile of the occupation $n(x,t)$  following the quench. 
Our findings indicate that, similar to the homogeneous configuration, the supersonic modes are visible in the 
density $n(x,t)$ or the current $j(x,t)$ profiles as well. Surprisingly, inside the regular light cone, at long enough times for the system to stabilize, 
 a non-equilibrium steady state is developing with a flowing current of a \emph{constant} magnitude, irrespective of the strength of the
 interaction, corresponding to a universal  conductance
%~\footnote{Except for Eq.~\eqref{eq:conductance}, where the units are used explicitly,   we shall use units corresponding to $e=\hbar=1$, in which case the universal conductance becomes $G_{\rm nH} = {1/2\pi} $} 
\begin{gather}
G =  {e^2\over h}\, , 
\label{eq:conductance}
\end{gather}
similar to the case of non-interacting spinless fermions. This sounds very counterintuitive since non-hermitian realm is
naturally associated with dissipation, although the system is isolated. 
What is even more surprising is that when we time evolve the same initial state with a \emph{hermitian} LL Hamiltonian, the long time conductance of this non-equilibrium setting is
$e^2/hK$ with $K$ the Luttinger liquid parameter~\cite{giamarchi}, which keeps track all interaction effects.

To verify our analytical results we investigate numerically the non-equilibrium dynamics 
in  non-Hermitian lattice model using the time evolving block decimation (TEBD) algorithm~\cite{Vidal.2003,Vidal.2004,Vidal.2007}, 
and we find perfect agreement with the bosonization results. 
%To complete the 
%analysis, we perform a Hermitian versus non-Hermitian comparative analysis of the occupation and for the entanglement
%entropy as well~\cite{Bastianello.2020}. 

\emph{Non-Hermitian quench protocol.}
We consider a global quench~\cite{Mitra.2018} in which the initial state is an inhomogeneous non-interacting 
LL~\cite{Mitra.2010} that displays a step-like local 
density distribution, similar to the one obtained by joining two semi-infinite systems that are initially kept at different chemical potentials. Here 
we consider a symmetric configuration with a positional dependent chemical potential of the form $\mu(x)=\mu_0\, \rm{sgn}(x)$. 

For such a system the ground state can be constructed exactly,
and in the bosonic language it corresponds to the ground state of the shifted quantum harmonic oscillator Hamiltonian~\cite{giamarchi}
\begin{equation}
H_{\rm inh}=\sum_{q\neq 0} \omega_q\big [ b_q^\dagger b_q -\lambda_{q}(b_q^\dagger+b_q)\big] ,
\label{eq:Hi}
\end{equation}
where $\omega_q = v|q|$ is the energy of the bosonic excitations, $b_q$ the annihilation operator for the density waves, and 
$\lambda_q = \mu_q/v\sqrt{2\pi|q| L}$ is a characteristic scale for the displacement of the ground state, with 
$\mu_q$, the Fourier transform of  the chemical potential profile $\mu(x)$.
The Hamiltonian~\eqref{eq:Hi} can be diagonalised exactly 
in terms of the shifted bosonic operators $a_q$, defined as  $a_q = b_q -\lambda_q$, which allows to
construct the GS as
\begin{equation}
\ket{\Psi_0}= \prod_q e^{-{|\lambda_q|^2 \over 2}} e^{-\lambda_q b_q^\dagger}\ket{0}\, ,
\label{eq:Psi_0}
\end{equation}
where $\ket{0}$ is the vacuum state for the $b_q$ operators, $b_q\ket{0}=0$. 
The  vacuum state $\ket{0}$ contains no bosonic excitations and represents the GS of the homogeneous setup.
 The ground-state
 $\ket{\Psi_0}$ represents the vacuum state for the $a_q$ operators,  $a_q\ket{\Psi_0} = 0$, and 
 by construction it is properly normalized, $\average{\Psi_0|\Psi_0} = 1$. 
At $t=0$, the Hamiltonian $H$, governing the evolution of the system, suddenly changes from $H_{\rm inh}$ to a non-Hermitian PT-symmetric 
Hamiltonian $H_{\rm nH}(t)$, i.e., $H_{\rm nH}(t) = H_{\rm inh} \Theta(-t) + H\Theta(t)$, 
by switching on the interaction and turning off the chemical potential, $\mu(x)= \mu(x)\Theta(-t)$. 
Following the quench, the evolution is governed by the Hamiltonian
\begin{equation}
H=\sum_{q\neq 0} \omega_q b_q^\dagger b_q
+\frac{ig_q}{2}\left[b_qb_{-q}+b_q^+b_{-q}^+\right].
\label{eq:H_nH}
\end{equation}
The Hamiltonian $H$ is similar to an interacting LL but with an imaginary interaction $ig_q$ instead.  
In general we shall use the parametrisation $g_q=g_2|q|$ to describe the strength of the interaction. 
Although the Hamiltonian is non-hermitian, as long as $g_2<v$, the energy spectrum of $H$ remains real as $\tilde v|q|$ with $\tilde v=\sqrt{v^2+g_2^2}$
and $H$ belongs to the PT-symmetrical models. For larger $g_2$, the system develops an instability~\cite{Dora.2020}.

\emph{Density and current profile.} 
Following the quench, the time evolution of the conventional~\cite{giamarchi} density and the current profiles are
\begin{gather}
\begin{bmatrix}
n (x,t) \\
j (x,t) 
\end{bmatrix}
 =-{1\over \pi {\cN}(t) } \bra{\Psi_0 (t)} \begin{bmatrix}
\partial_x \phi(x) \\
\partial_x \theta(x) 
\end{bmatrix}
  \ket{\Psi_0(t)} \, , 
\label{eq:n_nH}
\end{gather}
where $\ket{\Psi_0(t)}=e^{-i H t}\ket{\Psi_0}$ describes the  the non-unitary
time evolution on the initial wave function $\ket{\Psi_0}$ with Hamiltonian~\eqref{eq:H_nH}, and 
${\cN}(t)=\langle \Psi_0(t)|\Psi_0(t)\rangle$  is the norm of the wave function, while $\phi(x)$ and $\theta(x)$
are the regular LL fields, defined in terms of the $b_q$ operators~\cite{giamarchi, SM}. 

In general, for a hermitian Hamiltonian governing the dynamics, the
wave function is properly normalized to 1, i.e. ${\cN}(t)=1$, 
while in the non-Hermitian realm this is no longer true, and the norm explicitly depends on time.
Eq.~\eqref{eq:n_nH} can be used to find the initial  density profile as well, 
$n_{\rm inh}(x,t=0) = {1\over \pi v}\mu(x) + n_0 $, with $n_0$ representing the homogeneous background~\cite{SM},
indicating that the inhomogeneous contribution is determined exclusively by the shape of the chemical potential. 
At the same time, the initial current is zero. 

 Following the strategy putted forward in Ref.~\cite{Dora.2020}, we can evaluate $n(x,t)$
 and $j(x,t)$
using the pseudo-Heisenberg time evolution approach. Let us illustrate the derivation in terms of the density profile, 
the derivation for the current profile is similar with the proper replacement $\phi(x)\to \theta(x)$ in Eq.~\eqref{eq:n_nH}. 
We first introduce the pseudo-Heisenberg
time dependent fields $\psi(x,t>0) = e^{iH t}\psi(x)e^{-iH t}$ and the non-Hermitian forward and backward time 
evolution operator $U(t) = e^{i H ^\dagger t}e^{-iH t}$
in terms of which $n(x,t)$ becomes
\begin{equation}
n (x,t) =-{1\over \pi} \frac{\bra{\Psi_0} U(t) \partial_x \phi(x,t)\ket{\Psi_0}}{\average{\Psi_0|U(t)|\Psi_0}}.
\label{eq:n_xt}
\end{equation}
In any  typical Hermitian problem $U_{\rm H}(t)=1$, but since $[H,H^\dagger]\ne 0$, it follows that $U(t)\ne 1$ for the non-Hermitian evolution.  Notice that the time dependent fields are not constructed in the regular way as in the Heisenberg picture, 
but in a modified pseudo-Heisenberg way which is more suitable for our calculations. Using this 
construction,  $U(t)$ is rewritten as 
\begin{gather}
U(t) = \prod_{q>0} e^{C_+(q,t) K_+(q)}e^{C_0(q,t) K_0(q)}e^{C_-(q,t) K_-(q)},
\label{eq:U}
\end{gather}
in terms of the generators for the $SU(1,1)$ algebra
$
K_0(q) = {1\over 2}\big(b_q^\dagger b_q + b_{-q}b_{-q}^\dagger \big)$, 
$K_+(q) = b_q^\dagger b_{-q}^\dagger$, and $K_-(q) = b_{-q}b_q$.
%By matching the definition of the evolution operator with 
%the construction in Eq. ~\eqref{eq:U} it follows that the coefficients $C_{\pm, 0} (q,t)$ can be expressed
% in terms of the renormalized excitation energy $\toq=\tilde v|q|$, 
% and the interaction strength $g_q$.  
Using the construction in Eq.~\eqref{eq:U} and the  standard Baker-Hausdorff expressions,
 the time dependence of the norm of the wave function is evaluated as
\begin{equation}
{\cN}(t) = \prod_{q>0}\frac{\toq^2}{\toq^2-2g^2\sin^2\toq t}e^{2|\lambda_q|^2\frac{g \toq \sin 2\toq t+2g^2\sin^2\toq t}{\toq^2-2g^2\sin^2\toq t} }.
\label{eq:N}
\end{equation}
Details on the derivation of the Eq.~\eqref{eq:N} are discussed in \cite{SM}.
To evaluate the numerator in Eq.~\eqref{eq:n_xt} and calculate the time dependence of the density profile, it is required to know the time evolution of the $b_q(t)$ operators. Their time dependence  relies explicitly on the form of the final Hamiltonian and can be expressed exactly in terms
of two Bogoliubov coefficients $u_q(t)$ and $v_q(t)$ to which the time dependence is completely transferred.
In  evaluating $n(x,t)$ and $j(x,t)$ we normal order the product of various 
$b_q$ and $b_q^\dagger$ operators. Their  expressions are derived in Ref.~\cite{SM}, 
here we just present the final results. We can express them as a sum of two contributions, i.e.,
 $n(x,t) = n_{\rm r}(x,t) + n_{\rm s}(x,t)$ and  $j(x,t) = j_{\rm r}(x,t) + j_{\rm s}(x,t)$ 
where 
\begin{gather}
\begin{bmatrix}
n_{\rm r} (x,t) \\
j_{\rm r} (x,t) 
\end{bmatrix}
 = {1\over 2\pi}
 \begin{bmatrix}
 {1\over \tilde v} &  {1\over \tilde v}\\
1 & -1
\end{bmatrix}
\begin{bmatrix}
\mu(x -\tilde v t) \\
\mu(x +\tilde v t) 
\end{bmatrix}\, ,
\label{eq:nj_r}
\end{gather}
are the contributions  that provide the development of the regular light cone after the quench. 
The other contributions,  to  $n_{\rm s}(x,t)$  and $j_{\rm s}(x,t)$ 
are more involved and describe the  supersonic modes~\cite{SM}.  
For example, the supersonic current
is evaluated to
\begin{gather}
j_s(x,t) ={1\over\pi  L} \sum_{q>0}  e^{-\alpha|q|} (u_q(t)+v_q(t))\times\nonumber\\
{ 2 g^2 \sin^2 \toq t-u_p(t)\,g\,\toq\sin\toq t \over \toq^2 -2g^2\sin^2 \toq t } \cos (q x) \mu_q\, ,
\end{gather}
where the time dependent denominator is responsible for the supersonic modes and the multiple light cones.
Although it displays a strong spatial and temporal behaviour in general, it vanishes 
in the long time limit, $t\to \infty$, close to the center $x\sim 0$, implying that the only the 
regular contribution $j_r(x,t)$ controls the transport properties in the steady state. 
 
\emph{Post-quench conductance.} Following the quench, the initial domain wall develops into 
a central region, delimited by the boundaries of the lightcone. This regions extends 
ballistically in time in both directions with the light speed velocity $\pm \tilde v$. In a weak sense, in this region a local steady state is formed 
as the density equilibrates, but, due to the chemical potential drop, a particle current flows continuously. This allows us to define the conductance that characterise the non-equilibrium steady state  across the interface 
\begin{gather}
G =\frac 12  \frac{d\, j(x,t)}{d\,\mu_0}\bigg | _{\substack{x\to 0\\ \mu_0\to 0}}\, .
\label{eq:conductance_def}
\end{gather}
The regular contribution to the current gives a contribution to the conductance 
$G_{\rm r} =  G_0$ with $G_0=e^2/h$ the conductance quantum upon reinserting original units. 
Interestingly,  the anomalous contribution to the current from supersonic modes vanishes in the long time limit, implying $G_{\rm s}=0$, and the 
conductance acquires an  universal value given by Eq.~\eqref{eq:conductance} irrespective of the 
strength of the interaction. This  conductance is the unitary conductance of a single spinless channel and, at the same time, it corresponds  
to the conductance of a LL  connected to leads in equilibrium~\cite{safi,maslov}. 

This is to be contrasted to the post-quench conductance of a hermitian LL, characterised by  the Luttinger parameter~\cite{giamarchi} $K$,   
given by $G_{\rm H} = {G_0/K}$, thus in case of the unitary evolution the conductance strongly depends on the interaction strength~\cite{Mitra_b.2010} from the ensuing non-equilibrium state.
\begin{figure}[t!]
\includegraphics[width=0.49\columnwidth]{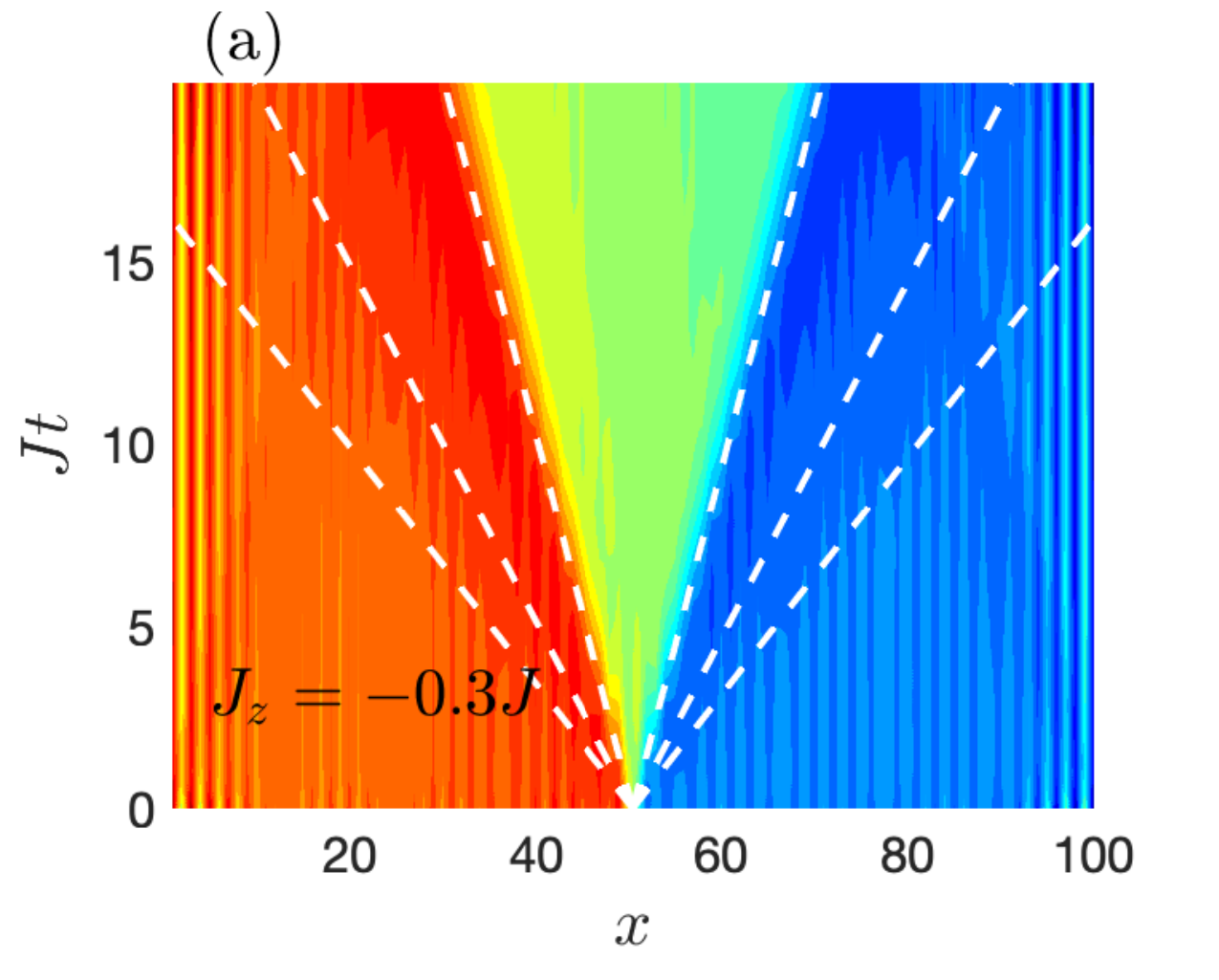}
\includegraphics[width=0.49\columnwidth]{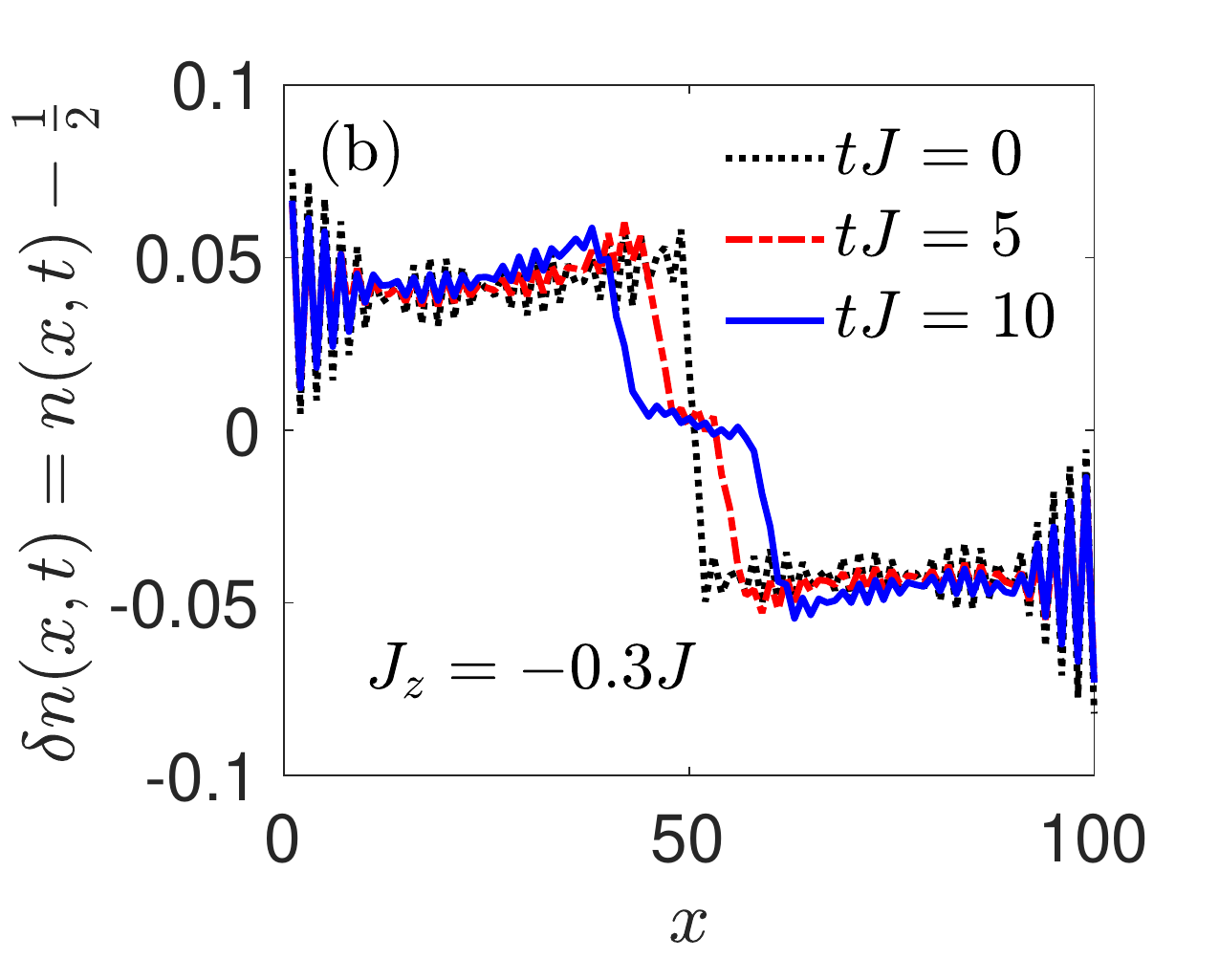}
\includegraphics[width=0.49\columnwidth]{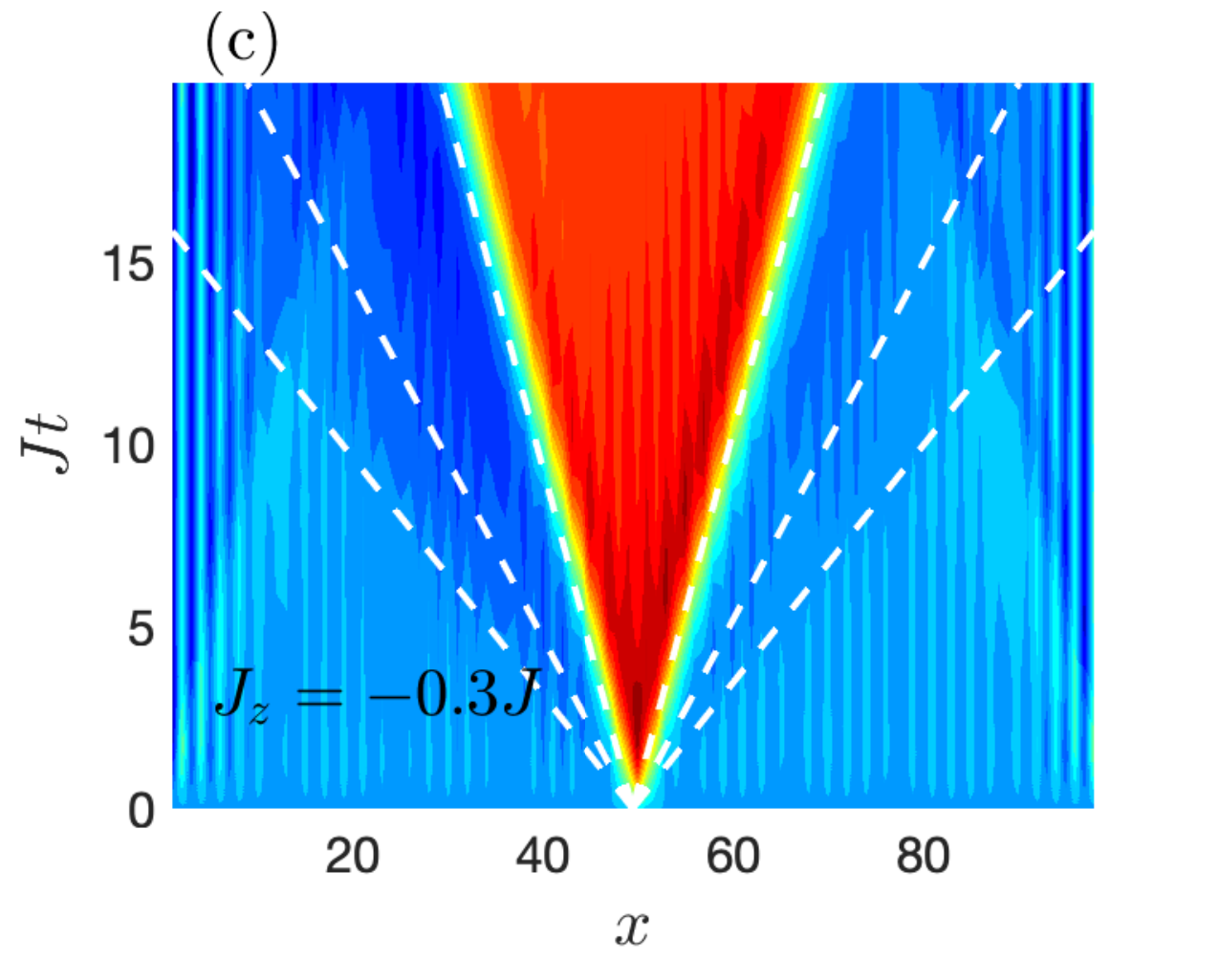}
\includegraphics[width=0.49\columnwidth]{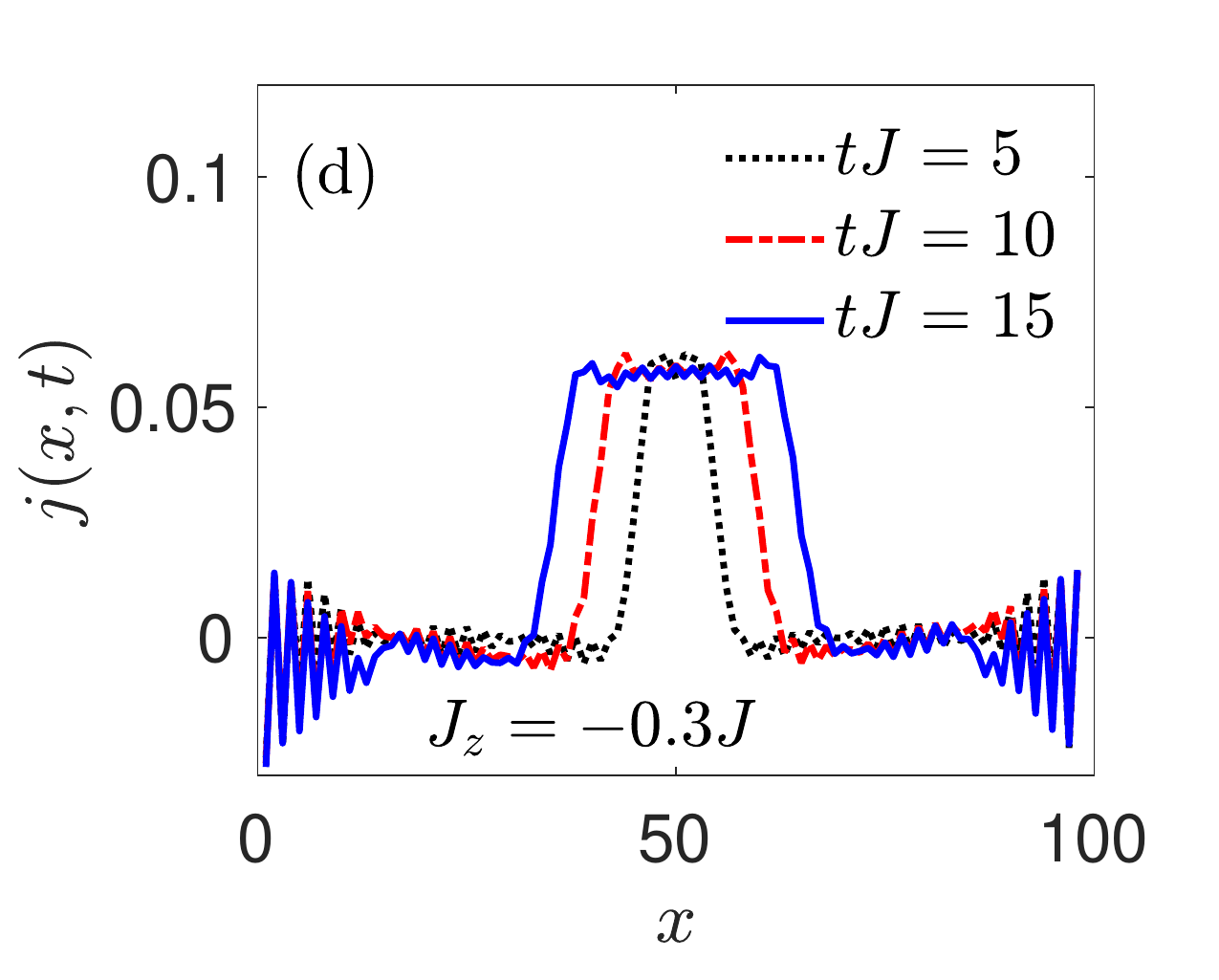}
\caption{  (a) Formation of the regular light cone as well as the supersonic modes 
in the density profile $n(x,t)$ for $J_z = -0.3J$. (b) Cuts at a
given time along the chain, displaying particles accumulation between the regular and supersonic light cones.
(c) Density plot for  the current  $j(x,t)$ (d) Evolution of the  current along the chain. In panel (a) and (c)
the white dashed lines represent the world lines for the light cones boundaries. 
 }
\label{fig:nH_evolution}
\end{figure}

\emph{Lattice model.} We corroborate our analytical results with a numerical analysis. For that we 
investigate numerically all the features that we have addressed so far, such as the formation of the light cone, the presence of the 
supersonic modes and most importantly we calculate the conductance across the interface following the quench in a one dimensional spinless lattice model. 
The initial state is constructed as a matrix product state 
 by performing density matrix renormalization group (DMRG) calculations~\cite{White.1992} on the non-interacting spinless Hamiltonian
\begin{gather}
H_{\rm inh}^{\rm (lat)}= \sum_{m = 1}^{N} \mu_m c^+_{m}c_m +\sum_{m=1}^N \frac{J}{2} \left(c^+_{m+1}c_m +\textmd{h.c.}\right),
\end{gather}
subject to a site dependent chemical potential of the form $\mu_m=\mu_0\,{\rm sgn}(m-N/2)$.  Here $c^\dagger_m$ are the creation operator at site $m$ along the chain. In
our calculations we fixed  the chain length to $N=100$,  while 
$J$, the nearest neighbour hopping, represents the energy unit. 

The MPS wave function is then  evolved in time 
 using the TEBD algorithm~\cite{Vidal.2004, Vidal.2007}, 
 with a non-hermitian evolution operator constructed from the 
Hamiltonian
\begin{gather}
H^{(\rm lat)}=\sum_{m=1}^N \frac{J+iJ_z}{2} \left(c^+_{m+1}c_m +\textmd{h.c.}\right)-i\frac{J_z\pi}{2} n_{m+1}n_m,
\label{eq:xxz}
\end{gather}
where $J_z$ is real and denotes the nearest neighbour interaction. This  model is not exactly PT-symmetric, but the low energy part of its spectrum can be considered
real, which influences the early time dynamics~\cite{Dora.2020}.
When $J_z$ is turned imaginary, $J_z\to -i J_z$, the model becomes 
the regular XXZ hermitian Heisenberg model, which is  Bethe-Ansatz solvable~\cite{giamarchi}  with a sound velocity  $v_{\rm H}\approx J+(1-\pi^2/8)J_z^2/J$. On the other hand, the low energy excitations for the non-Hermitian version are sound waves with sound velocity $\tilde v\approx J+(\pi^2/8-1)J_z^2/J$.
\begin{figure}[t!]
\includegraphics[width=0.7\columnwidth]{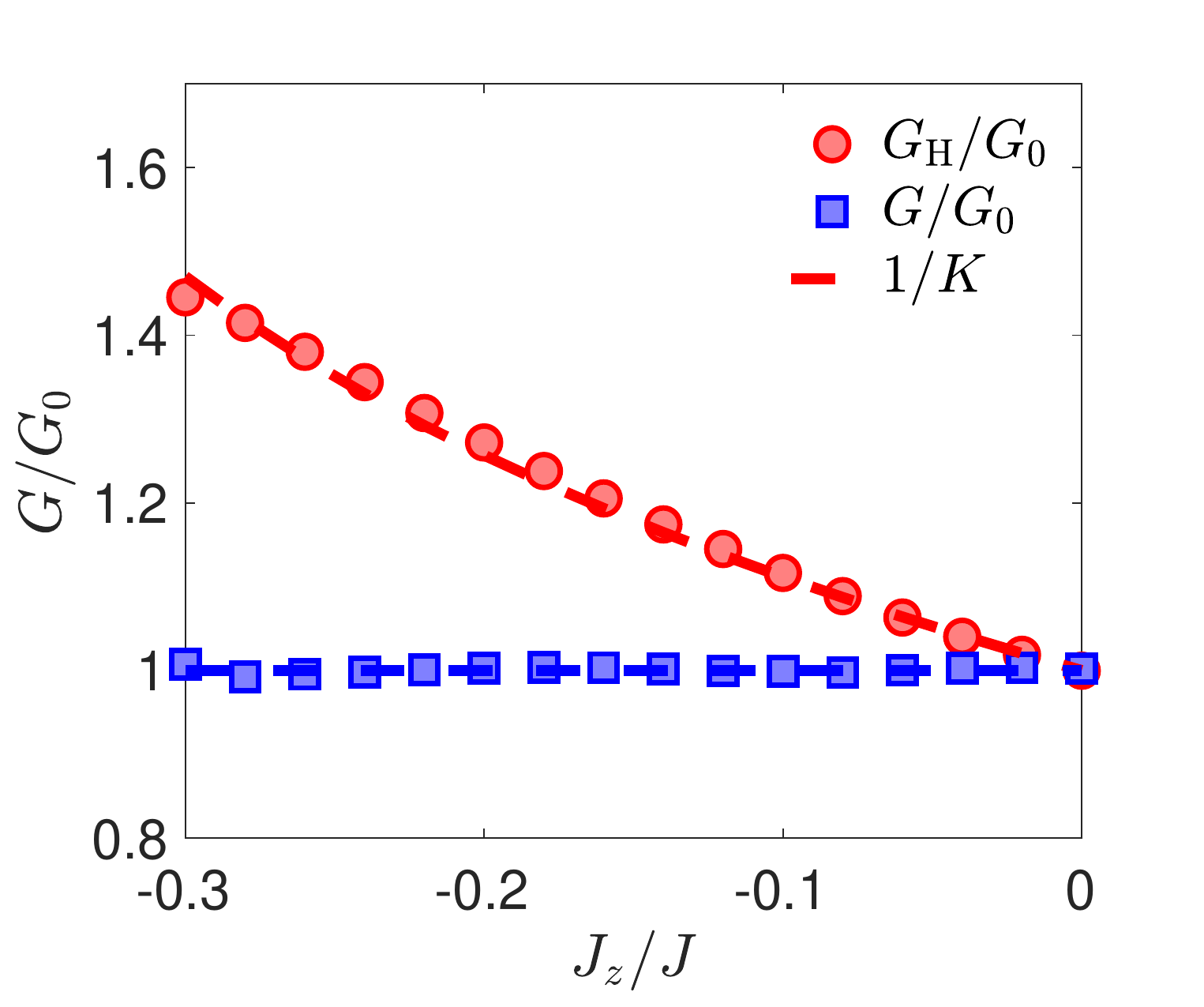}
\caption{ The blue square symbols represents the universal conductance of the PT non-hermitian Luttinger model indicating the universal behavior 
predicted in Eq.~\eqref{eq:conductance}.  The red symbols corresponds to  the conductance of the  Hermitian interacting LL  model, 
consisting in replacing $J_z\to -iJ_z$ in Eq.~\eqref{eq:xxz}.  The dashed line are fits with the analytical expressions displayed in the legend. 
\label{fig:conductance}
}
\end{figure}
Fig.~\ref{fig:nH_evolution}(a) displays the density plot for the occupation $n(x,t)$ along the chain as function of time. The regular light cone is clearly visible but also the formation of the second and third supersonic modes that propagates at velocities $v_n =n\tilde v$, $n=2,3$. 
The world lines for the boundaries of the lightcones are displayed with dashed lines.

The formation of various fronts is also visible in the density plot displaying the current $j(x,t)$ as well. Furthermore, 
 Fig.~\ref{fig:nH_evolution}(b,d) present several cuts at fixed times along the chain for the density  and current profiles. 
In the cuts representing $j(x,t)$ the formation of the plateaux inside the light cone, displaying the region with the constant currents  are clearly visible. 

We compute the conductance numerically by using Eq.~\eqref{eq:conductance_def}. 
For that, we find the stationary current at the interface in the long time limit, 
$tJ \sim 20 $, for various initial chemical potentials drops $\mu_0$ and then take the derivative numerically. 
We perform an average over a time interval of the order of 2 - 3$/J$ to remove 
the local oscillations sometimes visible in the current. For small enough $\mu_0$'s the stationary currents depend linearly on the $\mu_0$, which allows us to  extract the conductance as the slope of the current. The final result for the non-Hermitian conductance 
is displayed in Fig.~\ref{fig:conductance}. Irrespective of the value of the coupling strength $J_z$, the conductance remains universal and equal to the conductance quantum in perfect agreement with bosonization from Eq.~\eqref{eq:conductance}. 
We also display the results for the Hermitian evolution, by replacing $Jz\to -i J_z$ in Eq.~\eqref{eq:xxz}  where the analytical prediction from Ref.~\cite{Mitra_b.2010} fits also perfectly.

%
%\begin{figure}[h!]
%\includegraphics[width=0.49\columnwidth]{current_average_nh_Jz.pdf}
%\includegraphics[width=0.49\columnwidth]{current_average_nh_mu.pdf}
%\includegraphics[width=0.9\columnwidth]{conductance_nh.pdf}
%
%\caption{ (a) Non-hermitian evolution of the  current density as function of time for various $J_z$ and for $\mu = 0.15 J$. (b) The stationary current across the interface as function of the chemical potential difference. It displays a linear dependence for small $\mu$'s but saturates at large $\mu$ or the order of bandwidth. The slope of the curves 
%represent the conductance G.  (c) Non-hermitian conductance $G$ as function of the coupling $J_z$. 
%The horizontal line represents the conductance of a non-interacting LL chain corresponding to $G_0=K/\pi$, with $K=1$, the Luttinger parameter for non-interacting system. The conductance is always less than $K/\pi$. 
%\label{fig:current_h}
%}
%\end{figure}

\emph{Conclusions.}
We study the non-equilibrium dynamics and transport of a PT-symmetric Luttinger liquid when the model is quenched 
from a domain wall initial state. Due to non-unitary time evolution, we identify the formation of supersonic modes
on top of the regular light cone both in the density and  current profiles after the quench.
Most importantly, we find the universal value, $e^2/h$  for the 
conductance at the interface, which is a very robust analytical 
result, benchmarked by the numerical simulations. Moreover, for a unitary time evolution with a hermitian Luttinger liquid Hamiltonian, we resulting non-equilibrium conductivity gets
heavily renormalized by the interaction as $e^2/hK$.
Our setup can in principle be realized in dissipative lattices~\cite{Rauer.2016, Erne.2018} for which 
our predictions can be tested experimentally.

\emph{Acknowledgments.}
This research is supported by the National Research, Development and Innovation Office - NKFIH   
within the Quantum Technology National Excellence Program (Project No. 2017-1.2.1-NKP-2017-00001), K119442.

\bibliography{references}

\newpage

\section{Supplementary Material}

\subsection{Initial density }
The time and spatial evolution of the density profile $n(x,t)$ is captured by the Eq.~\eqref{eq:n_nH}
in the main paper. Using Eq.~\eqref{eq:n_nH} at $t=0$, we can calculate the
initial profile of the  local density. It can  be shown
 that it is modelled by the spatial dependence of the chemical potential  $\mu(x)$. 
To show that, we express the
 Luttinger field $\phi(x)$ in terms of the $a_q$ operators as $\phi(x) = \phi_a(x)+\delta\phi(x)$, where $\phi_a(x)$ has the usual form~\cite{giamarchi}
\begin{equation}
\phi_a(x)  = \phi_{\rm h}(x) -{i \pi\over L}\sum_{q\ne 0}\sqrt{L|q|\over 2\pi}{1\over q}e^{-i q x} (a_q^\dagger+a_{-q})\, ,
\end{equation}
where $\phi_{\rm h}(x)= -(N_L+N_R){\pi x\over L}$ represent the background local field   and $\delta\phi(x)$ is the inhomogeneity induced
by the chemical potential. 
\begin{equation}
\delta\phi(x) = {2 i \pi\over L}\sum_{q\ne 0}\sqrt{L|q|\over 2\pi}{1\over q}e^{-i q x}\lambda_q.
\label{eq:inh_contr}
\end{equation} 
The initial density profile can be evaluated as
\begin{equation}
n_{\rm i} (x) =-{1\over \pi}\bra{\Psi_0} \partial_x \phi(x)\ket{\Psi_0}\, , 
\end{equation}
with the expectation value taken  with respect to the ground state of $H_{\rm inh}$ given in Eq.~\eqref{eq:Hi}. Taking into account that 
$\ket{\Psi_0}$ is the vacuum state for the $a_q$ operators, 
it immediately implies that 
\begin{gather}
-{1\over \pi }\bra{\Psi_0} \partial_x \phi_a(x)\ket{\Psi_0}=-{1\over \pi} \bra{\Psi_0}  \partial_x \phi_{\rm h} (x)\ket{\Psi_0}=  n_0,
\end{gather} 
with $n_0 = (N_L+N_R)/L$ representing the homogeneous density background.  The inhomogeneous contribution in Eq.~\eqref{eq:inh_contr} is Fourier transformed
back to the real space, and the initial density profile is simply
\begin{equation}
n_{\rm inh}(x, t=0) = {1\over \pi v}\mu(x) + n_0\,.
\label{eq:n_i}
\end{equation}
Eq.~\eqref{eq:n_i} shows that the profile of the 
density in the initial state is determined exclusively by the chemical potential
and follows exactly its shape. Notice that in deriving Eq.~\eqref{eq:n_i} no 
particular shape for $\mu(x)$ has been considered so the result 
is valid for any spatial distribution of the chemical potential.

\subsection{Norm of the wave function}\label{sec:norm}
The evaluation of the norm ${\cN}(t)$ in Eq.~\eqref{eq:n_xt} is done by using the expression for the evolution operator $U(t)$
from Eq.~\eqref{eq:U}. Keeping in mind that $\lambda_q=-\lambda_{-q}$ is an antisymmetric function, the norm is cast in the following form
\begin{gather}
{\cN}(t) = \prod_{q>0}e^{-2|\lambda_q|^2}\bra{0} e^{-\lambda_q^*(b_q-b_{-q})} e^{C_+(q,t) K_+(q)}\nonumber\\
e^{C_0(q,t) K_0(q)}e^{C_-(q,t) K_-(q)}e^{-\lambda_q(b_q^\dagger-b_{-q}^\dagger)} \ket{0}\, .
\label{eq:N1}
\end{gather}
Here we need to compute the expectation value with respect to the initial state $\ket{\Psi_0}$ given in Eq.~\eqref{eq:Psi_0}.
In evaluating \eqref{eq:N1}, the strategy is to normal order the exponentials 
with the annihilation operators $b_q$ to the right and the creation operators $b_q^\dagger$ to the left. Let us discuss here the
reordering of  the last two exponents in~\eqref{eq:N1}, as the rest of the calculation is conceptually the same. For that, we first expand the last 
exponential into Taylor series
\begin{equation}
e^{-\lambda_q(b_q^\dagger-b_{-q}^\dagger)} = \sum_{n=1}^{\infty} {1\over n!}(-\lambda_p)^n(b_q^\dagger-b_{-q}^\dagger)^n\,.
\end{equation}
Next, keeping in mind that $K_-(q,t)=b_{-q}b_q$, it can be readily shown using the Baker-Hausdorff formula that 
\begin{eqnarray}
e^{C_-(q,t) K_-(q)} b_p^\dagger &=&(b_q^\dagger + C_-(q,t) b_{-q})e^{C_-(q,t) K_-(q)}\nonumber\\
e^{C_-(q,t) K_-(q)} b_{-p}^\dagger &=&(b_{-q}^\dagger + C_-(q,t) b_{q})e^{C_-(q,t) K_-(q)}\,.\nonumber
\end{eqnarray}
Performing the rotation for the whole series and reshaping the result back into an  exponential form we obtain
\begin{gather}
{\cN}(t) = \prod_{q>0}e^{(C_-(q,t)-2)|\lambda_q|^2} \bra{0} e^{-\lambda_q^*(b_q-b_{-q})} e^{C_+(q,t) K_+(q)}\nonumber\\
e^{C_0(q,t) K_0(q)}e^{-\lambda_q(b_q^\dagger-b_{-q}^\dagger)} \ket{0}\,.
\label{eq:N2}
\end{gather}
Performing similar transformations to fully normal order the expression we obtain for the norm factor
\begin{gather}
{\cN}(t) = \prod_{q>0}e^{C0(q,t)\over 2} e^{(C_+(q,t)+C_-(q,t)-2+2 e^{C_0(q,t)\over 2})|\lambda_q|^2} \,.
\label{eq:N3}
\end{gather}
The exact expressions for the coefficients $C_{\pm,0}(q,t)$ can be obtained 
following the strategy discussed in the main paper. The final expressions are
\begin{eqnarray}
C_0(q,t) &=& -2\ln\frac{\toq^2-2g^2\sin 2\toq t}{\toq^2} \nonumber \\
C_+(q,t) & =& \frac {g (i\oq\sin\toq t +\toq\cos\toq t)\sin \toq t}{\toq^2-2g^2\sin^2 \toq t} \\
C_-(q,t) &=& C_+^*(q,t)\, . \nonumber
\label{eq:C}
\end{eqnarray}
which allows us to  recover the expression for 
${\cN}(t)$ in Eq.~\eqref{eq:N}.
At $t=0$, obviously $N=1$ and the wave function is properly normalized.

\subsection{Time dependence of the evolution operators}\label{sec:b}
In this section we discuss the pseudo-Heisenberg time evolution of the annihilation/creation operators. Their time dependence is governed by 
$b_q(t)= e^{iH t}\,b_q\, e^{-iH t}$, which results in a Heisenberg equation of the form
\begin{gather}
\partial_t b_q(t) = i[H, b_q(t)], \phantom{aaa}
\partial_t b_q^\dagger(t) = i[H, b_q^\dagger(t)].
\end{gather}
Notice that the equation for $b_q^\dagger(t)$ is not recover from the one for $b_q(t)$ simply by hermitian conjugation. 
Computing  the commutators with the Hamiltonian~\eqref{eq:H_nH}, we obtain
\begin{gather}
\partial_t b_q(t) = -i\,\oq b_q(t)+g\, b_{-q}^\dagger (t)\nonumber \\
\partial_t b_{-q}^\dagger(t) = i\,\oq b_{-q}^\dagger(t) - g\, b_q (t)\,.
\label{eq:bq}
\end{gather}
To solve this set of equations we start by  searching for a general solution of the form
\begin{gather}
\begin{bmatrix}
b_q(t)\\
b_{-q}^\dagger(t)
\end{bmatrix}=
\begin{bmatrix}
u_q(t) & v_q(t)\\
- v_q^*(t) & u_q^*(t)
\end{bmatrix}
\begin{bmatrix}
b_q\\
b_{-q}^\dagger
\end{bmatrix}\,. \label{eq:uv}
\end{gather}
Such a solution is useful since  time dependence is  transferred to the Bogoliubov coefficients $u_q(t)$ and $v_q(t)$ entirely. Using the commutativity relation 
$[b_q, b_q^\dagger]=1$, it follows that $|u_q(t)|^2+|v_q(t)|^2=1$ and that they satisfy a differential equation similar to~\eqref{eq:bq}. 
Finally the solution is captured by the expressions 
\begin{gather}
u_q(t)=\cos \toq t-i{\oq\over \toq}\sin \toq t,\phantom{aaa}
v_q(t)={g\over \toq} \sin \toq t\,.
\end{gather}
in terms of the renormalized excitation energy of the quasiparticles.
 
\subsection{Density profile following the quench}
With the  exact analytical 
expressions for the coefficients  $u_q(t)$ and $v_q(t)$ at hand we can  
derive exact expressions for the time dependence of the density and current profiles 
following the quench by using Eq.~\eqref{eq:n_nH}
 Introducing the notation 
$\beta_q(x) = -{\pi\over L}e^{-iqx}e^{-\alpha |q|/2}\sqrt{L|q|\over 2\pi}$, 
for the overall prefactor in the $\theta(x)$ and $\phi(x)$ fields
and 
separating the contributions coming from the creation and annihilation operators 
we can write $\partial_x\phi(x,t) = \partial_x\phi_-(x,t)+\partial_x\phi_+(x,t)$, with 
\begin{eqnarray}
\partial_x\phi_+(x,t) = \sum_{q>0} \big(u_q^*(t)+v_q(t)\big)\big(\beta_q(x)b_q^\dagger +\beta_q^*(x)b_{-q}^\dagger\big)\nonumber\\
\partial_x\phi_-(x,t) = \sum_{q>0} \big(u_q(t)-v_q(t)\big)\big(\beta_q(x)b_{-q} +\beta_q^*(x)b_{q}\big)\nonumber\,.
\end{eqnarray}
In  evaluating $n(x,t)$ 
we follow the same strategy as the one in computing the norm ${\cN}(t)$, and normal ordering the product of various 
$b_q$ and $b_q^\dagger$ operators.

The piece $\partial_x\phi_-(x,t)$ gives a  partial contribution to the 
regular light cone. When  evaluating $n_-(x,t)$, the norm drops out
and we obtain
\begin{gather}
n_-(x,t) = {1\over 2\pi L \tilde v }\sum_{q\ne 0}\big( u_q(t)-v_q(t)\big)e^{-iqx}e^{-\alpha|q|/2}h_q\nonumber\\
\simeq {1\over 4\pi \tilde v}\big( \mu(x-\tilde v\, t)+\mu(x+\tilde v\, t) \big)
\end{gather}
The contribution from $\partial_x\phi_+(x,t)$ is more involved and contributes to the regular light cone as well as to the supersonic modes. Still, it can be brought into a compact expression of the form
\begin{gather}
n_+(x,t) = \frac{1}{2\pi L}\sum_{q>0 }\big (u_p^*(t)+v_p(t)\big )\times\nonumber\\ 
\frac{\toq^2-u_p(t)\toq g  \sin\toq t}{\toq^2-2g^2\sin^2\toq t} 2\sin(qx) e^{-\alpha|q|}\mu_q
\end{gather}
This part contributes to the regular light cone but also the supersonic modes.
To separate the two contributions  we expanded $n_+(x,t)$ in Taylor series in $\sin^2\toq t$. 
The zeroth order term contributes to the regular light cone while all the other higher order terms contributes to  the supersonic modes.
We then have $n(x,t) = n_{\rm r}(x,t) + n_{\rm s}(x,t)$
with
\begin{gather}
n_{\rm r}(x,t) ={1\over 2\pi \tilde v}\big( \mu(x-\tilde v t)+\mu(x+\tilde v t) \big)\,,
\label{eq:n_H}
\end{gather}
showing the development of the regular light cone after the quench and $n_s$ describing the supersonic modes 
\begin{gather}
n_{\rm s}(x,t) = {1\over 2\pi L \tilde v }\sum_{q> 0}\sum_{n=1}^{\infty}\big (u_p^*(t)+v_p(t)\big )\times\nonumber\\ 
\Big(1-u_p(t){g \sin \toq t\over \toq}\Big) \Big({\sqrt 2 g\sin \toq t\over \toq } \Big)^{2n} 2 i \sin (qx)e^{-\alpha|q|}\mu_q\, , 
\label{eq:n_perturb}
\end{gather}
and describes the supersonic modes. Notice that when the model is hermitian, all the terms 
$\sim \sin\toq t$ cancel and the contribution $n_{\rm s} (t)$ vanishes. 
For a given profile  of the initial chemical potential $\mu(x)$, 
the various integrals in Eq.~\eqref{eq:n_perturb} can be computed term by term in the perturbative expansion. 
If we consider the simplest chemical potential profile corresponding to a step-like function of the form
$\mu(x)=\mu_0\, \rm{sgn}(x)$, the corresponding Fourier transform is $\mu_p \propto \mu_0 {2 i \over p}$, 
and the integrals (sums) in Eq.~\eqref{eq:n_perturb} can be
 performed order by order  in the  
perturbation theory in $g_2/\tilde v$ to reveal to supersonic modes. 

%For example, keeping  the 
%contribution from the first Bogoliubov coefficient $u(x,t)$ we have up to the second order 
%\begin{gather}
%n_{\rm s}(x,t) \simeq -{v\over 2\tilde v^2}\Big(\mu(x - \tilde v t) + {v\over \tilde v}\mu(x+\tilde v t)\Big)\nonumber\\
% + {\tilde v+ 2v \over 2\tilde v^2} ( \mu(x-3 \tilde v t) +  \mu(x+3 \tilde v t )\Big)\Big( {g_2\over \tilde v} \Big )^2
% +{\cal O }\Big({g_2\over \tilde v}\Big)^3
% \label{eq:n_perturb}
%\end{gather}
%which besides the regular sonic modes, it also indicates the presence of another front propagating with $3 \tilde v$. 

\subsection{Current profile following the quench}
Once the chemical potential is turned off, and the interaction is quenched 
the initial domain wall induces a change in the local density and complementary
a current flow across the domain wall. Furthermore, the domain wall extends and transforms into
a central region inside the lightcone characterised by a steady current.
The current density can be evaluated in terms of the $\theta(x,t)$ field
\begin{equation}
\theta(x)  =  {i \pi\over L}\sum_{q\ne 0}\sqrt{L|q|\over 2\pi}{1\over |q|}e^{-i q x} \Big (b_q^\dagger - b_{-q}\Big )\, ,
\label{eq:theta}
\end{equation} 
as
\begin{equation}
j (x,t) =-{1\over \pi} \frac{\bra{\Psi_0 (t)} \partial_x \theta(x)\ket{\Psi_0(t)}}{{\cN}(t)}\, , 
%\label{eq:j_nH}
\end{equation}
which allows us to compute the current profile in a similar fashion as we computed the local density. Performing similar steps we can express the total current as a sum of the regular light cone contribution
\begin{gather}
j_{\rm r}(x,t) ={1\over 2\pi }\big( \mu(x-\tilde v t)-\mu(x+\tilde v t) \big)\,,
\label{eq:j_H}
\end{gather}
and a non-Hermitian part that can be again expressed as a power series of the form
\begin{gather}
j_{\rm s}(x,t) = {1\over 2\pi L}\sum_{q> 0}\sum_{n=1}^{\infty}\big (u_p(t)+v^*_p(t)\big )\times\nonumber\\ 
\Big(1-u_p(t){g \sin \toq t\over \toq}\Big) \Big({\sqrt 2 g\sin \toq t\over \toq} \Big)^{2n} 2  \cos(qx) e^{-\alpha|q|}\mu_q\, , 
\label{eq:j_nH}
\end{gather}
an expression that allows us to compute the current order by order in the perturbation theory in a manner similar to
Eq.~\eqref{eq:n_perturb}. 

%Notice that the hermitian components for the density and current given in Eqs.~\eqref{eq:n_H} and \eqref{eq:j_H}
%are consistent with the continuity equation. However, in the presence of dissipation, as simulated by the non-Hermitian Hamiltonian, the continuity equation is no longer satisfied. 

\end{document}